Freedom in Expansion and Asymptotic Integrability of Perturbed Evolution Equations

Yair Zarmi

Jacob Blaustein Institutes for Desert Research

Ben-Gurion University of the Negev, Sede-Boqer Campus

Midreshet Ben-Gurion, 84990 Israel

Abstract

It is shown that the emergence of obstacles to asymptotic integrability in the analysis of perturbed evolution equations may, often, be a consequence of the manner, in which the freedom in the expansion is exploited in the derivation of the equations. Algorithms exist, which yield perturbed evolution equations that are devoid of the obstacles for cases, in which, traditionally, obstacles are encountered. The derivation of the perturbed KdV equation for two physical systems (propagation of small amplitude disturbances on a shallow fluid layer, and the ion acoustic wave equations in Plasma Physics), where a second-order obstacle is anticipated, and of the Burgers equation (one-dimensional propagation of weak shock waves in an ideal gas), where a first-order obstacle is anticipated, is examined. In all cases, the anticipated obstacles to integrability can be avoided.

Integrable evolution equations provide approximate descriptions for complex dynamical systems. The properties of such equations enable one to obtain a wealth of information about solutions of the original systems [1-15]. When the terms that have been omitted in the derivation of an evolution equation are reinstated, they constitute a small perturbation. Often, the perturbed evolution equation is not integrable, either rigorously or in an order-by-order perturbative analysis (i.e., even "asymptotic" integrability is lost). From some order onwards, the perturbation contains, or generates in the dynamical equations, terms that have the capacity to spoil integrability. The existence of obstacles to asymptotic integrability has been demonstrated in the cases of the perturbed Burgers [16,17], KdV [18-20] and NLS [21, 22] equations.

Ways to handle the effect of obstacles on the solutions of perturbed evolution equations (based on different manners of exploitation of the freedom inherent in the expansion procedure) have been discussed in the literature in the cases of the perturbed KdV [18-20, 23-25], NLS [21, 22, 26] and Burgers [24, 27] equations. In this note, it will be shown that, often, the emergence of obstacles in a perturbed evolution equation may be a consequence of the expansion algorithm employed in the derivation of the equation. Moreover, there are algorithms, which generate perturbed evolution equations that are devoid of obstacles. The derivation of the KdV equation for two physical systems (one-dimensional propagation of small amplitude disturbances on a shallow fluid layer, and the ion acoustic wave equations in Plasma Physics) and of the Burgers equation (one-dimensional propagation of weak shock waves in an ideal gas) will be examined.

Let us begin with the perturbed KdV equation. Its generic form is [18-20]:

$$\begin{aligned} w_t = {} & 6ww_x + w_{xxx} + \varepsilon\left(30\alpha_1 w^2 w_x + 10\alpha_2 w w_{xxx} + 20\alpha_3 w_x w_{xx} + \alpha_4 w_{5x}\right) \\ & + \varepsilon^2 \begin{pmatrix} 140\beta_1 w^3 w_x + 70\beta_2 w^2 w_{xxx} + 280\beta_3 w w_x w_{xx} \\ +14\beta_4 w w_{5x} + 70\beta_5 {w_x}^3 + 42\beta_6 w_x w_{4x} + 70\beta_7 w_{xx} w_{xxx} + \beta_8 w_{7x} \end{pmatrix} + O\left(\varepsilon^3\right) \; . \qquad (1) \\ & \left(|\varepsilon| \ll 1\right) \end{aligned}$$

The coefficients in the perturbation depend on the physical problem. One needs to go to second order, because an obstacle to integrability is first encountered in that order. (The numerical coefficients in Eq. (1) are chosen so that if the $\alpha$ 's are all equal and the $\beta$ 's are all equal, the first- and second-order perturbations are proportional to symmetries of the KdV equation.)

To obtain an approximate solution, one expands $w$ in a near identity transformation (NIT):

$$w = u + \varepsilon u^{(1)} + \varepsilon^2 u^{(2)} + O\left(\varepsilon^3\right) \quad , \tag{2}$$

and expects the zero-order approximation, $u$, to obey a Normal Form (NF), which is constructed out of symmetries [18-20, 28-36]:

$$\begin{gathered} u_t = S_2[u] + \varepsilon \alpha_4 S_3[u] + \varepsilon^2 \beta_8 S_4[u] + O\left(\varepsilon^3\right) \\ \left( \begin{gathered} S_2 = 6 u u_x + u_{xxx} \\ S_3 = 30 u^2 u_x + 10 u u_{xxx} + 20 u_x u_{xx} + u_{5x} \\ S_4 = 140 u^3 u_x + 70 u u_{xxx} + 280 u u_x u_{xx} + 14 u u_{5x} + 70 u_x^3 + 42 u_x u_{4x} + 70 u_{xx} u_{xxx} + u_{7x} \end{gathered} \right) \end{gathered} \quad . \tag{3}$$

It is desirable that such a scheme be realizable for two reasons. First, Eq. (3) is integrable, and has the same single- and multiple-soliton solutions as the unperturbed equation. The only modification is in the dispersion relation, which determines the dependence of the soliton velocity on the soliton wave number. Second, one can ensure that the higher-order corrections are bounded.

In the NF analysis, it is customary to assume that the higher-order corrections, $u^{(n)}$, are differential polynomials in the zero-order solution, $u$ [16-22]. One then finds that an algebraic impasse is encountered in the second-order analysis: The computation of $u^{(2)}$ cannot be carried out unless the coefficients of the perturbation obey the relation [18-20]

$$5\left(3\alpha_1\alpha_2+4\alpha_2{}^2-18\alpha_1\alpha_3+60\alpha_2\alpha_3-24\alpha_3{}^2+18\alpha_1\alpha_4-67\alpha_2\alpha_4+24\alpha_4{}^2\right) \\ +21\left(3\beta_1-4\beta_2-18\beta_3+17\beta_4+12\beta_5-18\beta_6+12\beta_7-4\beta_8\right)=0 \quad , \tag{4}$$

Two ways out of this situation have been discussed in the literature, exploiting the freedom in the expansion in different manners. In the first way [18-20], $R[u]$, the obstacle to asymptotic integrability contained in the perturbation, is moved into the NF. Eq. (3) is then modified into

$$u_t=S_2[u]+\varepsilon\alpha_4 S_3[u]+\varepsilon^2\left(\beta_8 S_4[u]+R[u]\right)+O\left(\varepsilon^3\right) \ . \tag{5}$$

The advantage of this approach is that, given a bounded zero-order solution, $u$, the higher-order corrections are ensured to be bounded, as they are differential polynomials in $u$. The loss is that Eq. (5) is not integrable, as $R[u]$ is not a symmetry. As a result, the zero-order approximation, $u$, which solves Eq. (5), does not have the simple single- or multiple-wave structure of the unperturbed solution, and may not have a closed-form expression; it may have to be found numerically.

In the second way [23-25], one allows $u^{(2)}$ of Eq. (2) to be comprised of a differential polynomial in $u$, *plus* a non-polynomial term. This enables one to account for the obstacle, $R[u]$, by the non-polynomial component, and evades the need to shift the obstacle into the NF. The gain is that the NF retains its form of Eq. (3), hence, remains asymptotically integrable, and generates a zero-order solution that has the same structure as the unperturbed solution. The drawback is that the higher-order corrections contain non-polynomial terms, for which closed-form expressions are not known in general, and may have to be computed numerically. More important, these terms are not automatically guaranteed to be bounded as functions of $x$ and $t$. In fact, the freedom in the expansion must be invoked in a specific manner to ensure that they are bounded.

Let us return to the expansion procedure, which leads to Eq. (1), beginning with the problem of the propagation of small disturbances on the surface of a shallow layer of an incompressible and inviscid fluid, originally analyzed in [1]. In dimensionless quantities, the equations governing the two-dimensional irrotational flow are (we follow the notation of [37]):

$$\Phi_{zz} + \varepsilon \Phi_{\xi\xi} = 0; \qquad \Phi_z = 0 \qquad \text{on } z = 0 \quad , \tag{6}$$

$$\zeta - \Phi_\xi + \varepsilon \Phi_\tau + \tfrac{1}{2}\left(\Phi_z{}^2 + \varepsilon \Phi_\xi{}^2\right) = 0 \quad \text{on } z = 1 + \varepsilon\zeta \quad , \tag{7}$$

$$\Phi_z = \varepsilon\left(-\zeta_\xi + \varepsilon\zeta_\tau + \varepsilon\Phi_\xi\zeta_\xi\right) \qquad on\ z = 1 + \varepsilon\zeta \quad . \tag{8}$$

Here $\Phi(\tau, \xi, z)$ is the velocity potential, $z$ is the vertical coordinate, $\xi$ is the horizontal coordinate, $\tau$ is the time variable, the depth of the quiet fluid layer is equal 1, and the height of the disturbance above the top surface of the fluid is $\zeta(\tau, \xi)$. Finally, $\varepsilon \ll 1$ is the ratio of disturbance amplitude over layer depth. Eq. (6) is the equation of continuity of fluid dynamics, while Eqs. (7) and (8) are the conditions on the motion at the top surface.

To obtain the KdV equation with a perturbation through second order, one has to expand the problem through $O(\varepsilon^4)$. To this end, one writes

$$\Phi(\tau,\xi,z) = \sum_{n\geq 0} \varepsilon^n \Phi_n(\tau,\xi,z), \qquad \zeta(\tau,\xi) = \sum_{n\geq 0} \varepsilon^n \zeta_n(\tau,\xi) \quad . \tag{9}$$

One now uses Eq. (6) to successively solve for $\Phi_n$. The solution of each will contain a free function, $\theta_n(\tau, \xi)$. For instance,

$$\Phi_0(\tau,\xi,z) = \theta_0(\tau,\xi), \quad \Phi_1(\tau,\xi,z) = \theta_1(\tau,\xi) - \tfrac{1}{2}z^2\theta_{0,\xi\xi} \quad . \tag{10}$$

One then uses Eq. (7) to successively solve for the free functions $\theta_n(\tau, \xi)$. For instance,

$$\theta_0(\tau,\xi) = \partial_\xi^{-1}\zeta_0(\tau,\xi), \quad \theta_1(\tau,\xi) = \partial_\xi^{-1}\left(\zeta 1(\tau,\xi)\right) + \partial_\tau\partial_\xi^{-2}\zeta_0(\tau,\xi) + \tfrac{1}{2}\partial_\xi^{-1}\left(\zeta_0(\tau,\xi)^2\right) + \tfrac{1}{2}\partial_\xi\zeta_0(\tau,\xi) \ . \tag{11}$$

Having solved for $\theta_n(\tau, \xi)$, Eq. (8) becomes a differential equation for $\zeta_0(\tau, \xi)$, which through second order in $\varepsilon$, contains $\zeta_1(\tau, \xi)$, $\zeta_2(\tau, \xi)$ and $\zeta_3(\tau, \xi)$.

The next stage, which converts Eq. (8) to the KdV equation in a canonical form plus a perturbation, is rescaling according to:

$$\zeta_n(\tau,\xi) = \tfrac{2}{3} u_n(t,x), \quad \left(t = -(\tau/6) \quad , \quad x = \xi\right) \quad . \tag{12}$$

One is tempted to make a "simple" choice for $u_n(t, x)$, $n \geq 1$. An obvious choice is $u_n(t, x) = 0$. However, as will become clear later on, this leaves a perturbed KdV equation that contains a second-order obstacle to integrability. Another obvious choice is to require that Eq. (8) is obeyed separately order-by-order, and thereby successively solve for $u_n(t, x)$, $n \geq 1$. The equation for the zero-order approximation, $u_0$, becomes the unperturbed KdV equation. However, generation of unbounded contributions in the higher-order corrections can occur. For instance, the first-order term in Eq. (8) now obtains the form

$$u_{1,t} - 6\left(u_0\, u_1\right)_x - u_{1,xxx} = -u_0^{\,2}\, u_{0,x} + \tfrac{5}{3} u_0\, u_{0,xxx} + \tfrac{23}{6} u_{0,x}\, u_{0,xx} + \tfrac{19}{60} u_{0,xxxxx} \quad . \tag{13}$$

The right-hand-side of Eq. (13) contains the symmetry $S_3$ (see Eq. (3)), which resonates with the homogeneous part of the equation, hence, will generate an unbounded contribution in $u_1$. The symmetry can be incorporated in the equation for $u_0$ in a manner that ensures an integrable equation [10,16-22], hence may be subtracted from Eq. (13). The resulting equation for $u_1$ becomes:

$$u_{1,t} - 6\left(u_0\, u_1\right)_x - u_{1,xxx} = -\tfrac{21}{2} u_0^{\,2}\, u_{0,x} - \tfrac{3}{2} u_0\, u_{0,xxx} - \tfrac{5}{2} u_{0,x}\, u_{0,xx} \quad . \tag{14}$$

One needs to make a choice that will guarantee that $u_n$ are bounded. Aiming at an equation for $u_0$ that will generate a bounded solution, it is natural to assume that $u_n$, $n \geq 1$, are differential polynomials in $u_0$. The formalism allows for local as well as non-local terms (containing, e.g., $\partial_x^{-1} u_0$).

However, the latter turn out to unnecessarily complicate the resulting expressions. Excluding non-local terms, the allowed forms for $u_n$ are [18-20]:

$$\begin{aligned} u_1(t,\xi) &= a_1 u_0^{\,2} + a_2 \partial_x^{\,2} u_0 \\ u_2(t,\xi) &= b_1 u_0^{\,3} + b_2 u_0 \partial_x^{\,2} u_0 + b_3 (\partial_x u_0)^2 + b_4 \partial_x^{\,4} u_0 \\ u_3(t,\xi) &= c_1 u_0^{\,4} + c_2 u_0^{\,2} \partial_x^{\,2} u_0 + c_3 u_0 (\partial_x u_0)^2 + c_4 u_0 \partial_x^{\,4} u_0 + c_5 \partial_x u_0 \partial_\xi^{\,3} u_0 + c_6 (\partial_x^{\,2} u_0)^2 \end{aligned} \quad . \quad (15)$$

It is easy to check that $u_1$, of Eq. (15) does not satisfy Eq. (14). (In this order, it suffices to use the unperturbed KdV equation for $u_0$.) Thus, the attempt to solve for $u_n$, $n \ge 1$, successively fails, at least when $u_n$ are assumed to be differential polynomials.

The only option left, is to assume for $u_n$, $n \ge 1$, known forms and convert them into parts of the driving term in the equation for $u_0$. To this end, one uses Eq. (15), augmented by Eq. (12). Eq. (8) then obtains the form of Eq. (1). As an example, here are the values of the first few coefficients:

$$\begin{aligned} &\alpha_1 = \tfrac{1}{5} a_1 - \tfrac{1}{30}, \quad \alpha_2 = \tfrac{1}{6}, \quad \alpha_3 = \tfrac{3}{10} a_1 - \tfrac{3}{5} a_2 + \tfrac{23}{120}, \quad \alpha_4 = \tfrac{19}{60} \\ &\beta_1 = -\tfrac{1}{210} a_1 - \tfrac{1}{35} a_1^{\,2} + \tfrac{1}{20} b_1 - \tfrac{3}{35} c_1 + \tfrac{1}{630}, \quad \beta_2 = -\tfrac{13}{210} a_1 - \tfrac{1}{35} a_1 a_2 - \tfrac{3}{70} b_1 + \tfrac{1}{70} b_2 - \tfrac{3}{70} c_2 + \tfrac{4}{315} \end{aligned} \quad . (16)$$

The condition for the absence of a second-order obstacle to integrability, Eq. (4), becomes:

$$\begin{aligned} &\tfrac{1}{40} a_1 + \tfrac{171}{20} a_2 - \tfrac{9}{5} a_1^{\,2} + \tfrac{39}{5} a_1 a_2 - \tfrac{36}{5} a_2^{\,2} \\ &+ \tfrac{81}{10} b_1 - \tfrac{84}{5} b_2 - \tfrac{27}{10} b_3 + \tfrac{201}{2} b_4 - \tfrac{27}{5} c_1 + \tfrac{117}{10} c_2 - \tfrac{27}{10} c_3 - \tfrac{99}{2} c_4 + \tfrac{81}{5} c_5 - \tfrac{108}{5} c_6 - \tfrac{57}{80} = 0 \end{aligned} \quad , \quad (17)$$

which is easily obeyed by a wealth of choices for the free coefficients. (Clearly, if one chooses all the $u_n$ to vanish, the obstacle to integrability cannot be eliminated.)

Thus, exploiting the freedom in the expansion, it is possible to derive the perturbed KdV equation so that it is devoid of the second-order obstacle to integrability. Moreover, it is possible to assign values for eight of the coefficients (e.g., $b_3$, $b_4$, $c_1 - c_6$) such that *all* $\beta_n = 0$, so that the second-order perturbation vanishes identically! Eq. (4) then degenerates into

$$-4a_1 + 156a_2 + 2160a_1a_2 - 648a_1^2 - 1728a_2^2 + 9 = 0 \quad , \tag{18}$$

still, enabling the elimination of the second-order obstacle to asymptotic integrability.

Next, consider the case of the ion acoustic wave equations in Plasma Physics, governing the behavior of a collisionless plasma of cold ions and warm electrons in (1 + 1) dimensions [38, 7]. In terms of dimensionless quantities, the equations read [38, 39]:

$$n_\tau + \left(n\,v\right)_\xi = 0 \quad , \tag{19}$$

$$v_\tau + \left(\tfrac{1}{2}v^2 + \varphi\right)_\xi = 0 \quad , \tag{20}$$

$$\varphi_{\xi\xi} = e^\varphi - n \quad . \tag{21}$$

Eqs. (19) and (20) are the continuity- and momentum-conservation equations of fluid dynamics, respectively, and Eq. (21) is the Poisson equation for the electrostatic potential, $\varphi(\tau, \xi)$. $n(\tau, \xi)$ and $v(\tau, \xi)$ are the ion density and velocity, respectively, and the electron density is $e^\varphi$.

The first stage is rescaling according to:

$$n(\tau,\xi) = N(\sigma,x), \quad v(\tau,\xi) = V(\sigma,x), \quad \varphi(\tau,\xi) = \Phi(\sigma,x) \quad \left(\sigma = \varepsilon^3\tau\,, \quad x = \varepsilon\xi\right) \quad . \tag{22}$$

This leads to the natural small parameter, $\mu = \varepsilon^2$, in terms of which Eqs. (19) – (21) become:

$$\mu N_\sigma + \left(NV\right)_x = 0 \quad , \tag{23}$$

$$\mu V_\sigma + V\,V_x + \Phi_x = 0 \quad , \tag{24}$$

$$\mu \Phi_{xx} = e^{\Phi} - N \quad . \tag{25}$$

The stationary values of the solutions of Eqs. (24) – (25) are

$$N = 1 \quad , \quad V = \pm 1 \quad , \quad \Phi = 0 \quad . \tag{26}$$

As the results of the analysis do not depend on which of the two values of $V$ in Eq. (26) is chosen, $V = +1$ is adopted in the following. To derive the perturbed KdV equation through second order in $\mu$, one has to expand the functions and the equations through $O(\mu^4)$. Thus, we write

$$N = 1 + \sum_{n \geq 0} \mu^{n+1} N_n \quad , \quad V = 1 + \sum_{n \geq 0} \mu^{n+1} V_n \quad , \quad \Phi = \sum_{n \geq 0} \mu^{n+1} \Phi_n \quad . \tag{27}$$

The procedure delineated in the case of the shallow-water problem is now repeated. One uses Eq. (23) to successively solve for $N_n$, and then, Eq. (24) to successively solve for $\Phi_n$. Eq. (25) then becomes a differential equation for $V_0(\sigma, x)$, which, through second order in $\mu$, contains $V_1(\sigma, x)$, $V_2(\sigma, x)$ and $V_3(\sigma, x)$. To obtain the KdV equation in a canonical form one rescales according to:

$$V_n(\sigma, x) = -3 u_n(t, x), \qquad (t = 2\sigma) \quad . \tag{28}$$

One now assumes for $u_n$, $n = 1, 2, 3$ the differential polynomials in terms of $u_0$, as in Eq. (15). Eq. (25) now becomes Eq. (1) for $u_0$, with coefficients, a few of which are given below:

$$\begin{aligned} &\alpha_1 = -a_1 \,, \quad \alpha_2 = -\tfrac{3}{5} a_1 - \tfrac{9}{5} a_2 - \tfrac{3}{10} \,, \quad \alpha_3 = \tfrac{3}{10} a_1 - \tfrac{33}{10} a_2 + \tfrac{3}{40} \,, \quad \alpha_4 = -3 a_2 + \tfrac{3}{4} \\ &\beta_1 = \tfrac{9}{5} a_1^2 - \tfrac{12}{35} b_1 \,, \quad \beta_2 = \tfrac{3}{14} a_1 + \tfrac{24}{35} a_1^2 + \tfrac{27}{7} a_1 a_2 - \tfrac{9}{70} b_1 - \tfrac{9}{35} b_2 + \tfrac{9}{140} \\ &\beta_3 = -\tfrac{3}{35} a_1 - \tfrac{6}{35} a_1^2 + \tfrac{9}{2} a_1 a_2 + \tfrac{9}{140} b_1 - \tfrac{3}{10} b_2 - \tfrac{9}{70} b_3 - \tfrac{9}{560} \end{aligned} \quad . \tag{29}$$

Unlike the shallow-water problem, there is not enough freedom here to make the second-order perturbation vanish identically. However, Eq. (4) can be satisfied, so that the second-order obstacle to integrability is eliminated. Concurrently, a choice exists, for which

$$\beta_2 + \beta_3 = 2\beta_5 \ , \quad \left( b_3 = \tfrac{5}{4} a_1 + 3a_1^{\ 2} - 5\, a_1\, a_2 - \tfrac{1}{2} b_1 + b_2 \right) \ , \tag{30}$$

ensuring that the second-order perturbation can be expressed as a complete differential with respect to $x$. Thus, through second order, the resulting perturbed KdV equation can be written as a "continuity" equation:

$$u_{0,t} = \partial_x \left\{ 3u_0^{\ 2} + u_{0,xx} + \mu F_1 + \mu^2 F_2 \right\} \ . \tag{31}$$

In Eq. (31), $F_1$ and $F_2$ are the "fluxes", the divergence of which equals the first- and second-order perturbations, respectively.

In the case of the perturbed Burgers equation, an obstacle to integrability is encountered in the first-order perturbation. The generic form of the equation is [16, 17]

$$w_t = 2w w_x + w_{xx} + \varepsilon \left( 3\alpha_1 w^2 w_x + 3\alpha_2 w w_{xxx} + 3\alpha_3 w_x^{\ 2} + \alpha_4 w_{xxx} \right) + O\left(\varepsilon^2\right). \tag{32}$$

The obstacle to integrability emerges unless the coefficients obey the constraint [16, 17]:

$$2\alpha_1 - \alpha_2 - 2\alpha_3 + \alpha_4 = 0 \ . \tag{33}$$

Consider the derivation of Eq. (32) from the continuity and momentum-conservation equations in the case of a one-dimensional ideal gas [7, 17]:

$$\rho_\tau + \left(\rho v\right)_\xi = 0 \ , \tag{34}$$

$$\left(\rho v\right)_\tau + \left(\rho v^2 + P - \mu u_\xi\right)_\xi = 0 \, , \quad \left( P = \left(c^2 \rho_0/\gamma\right)\left(\rho/\rho_0\right)^\gamma \right). \tag{35}$$

In Eqs. (34) and (35), $\rho$ is the gas density, $v$–its velocity, $P$ is the pressure, $\mu$ is the viscosity coefficient, $\rho_0$ is the density of the quiet gas, $c$ is the speed of sound and $\gamma = (c_p/c_v)$ is the ratio of specific heats. One transforms and rescales variables according to

$$u(\tau,\xi) = \tilde{u}(\sigma,\chi), \quad \rho(\tau,\xi) = \tilde{\rho}(\sigma,\chi) \quad ,\left(\sigma = \varepsilon^2 \tau, \quad \chi = \varepsilon(\xi - c\tau)\right) \ . \tag{36}$$

Eqs. (34) and (35) are converted into

$$\varepsilon \tilde{\rho}_\sigma + (\tilde{\rho}\tilde{u})_\chi - c\tilde{\rho}_\chi = 0 \ , \tag{37}$$

$$\varepsilon\left\{(\tilde{\rho}\tilde{u})_\sigma - \mu \tilde{u}_{\chi\chi}\right\} + (\tilde{\rho}\tilde{u}^2)_\chi - c(\tilde{\rho}\tilde{u})_\chi + c^2(\tilde{\rho}/\rho_0)^{\gamma-1}\tilde{\rho}_\chi = 0 \ , \tag{38}$$

respectively. One now expands $\tilde{u}$ and $\tilde{\rho}$ in power series through $O(\varepsilon^3)$:

$$\tilde{u}(\sigma,\chi) = \varepsilon u_0 + \varepsilon^2 u_1 + \varepsilon^3 u_2 + \ldots, \quad \tilde{\rho}(\sigma,\chi) = \varepsilon \tilde{\rho}_0 + \varepsilon^2 \tilde{\rho}_1 + \varepsilon^3 \tilde{\rho}_2 + \ldots \ . \tag{39}$$

One then uses Eq. (37) to successively solve for $\tilde{\rho}_n$, to obtain

$$\begin{aligned}
&\tilde{\rho}_0(\sigma,\chi) = \frac{\rho_0}{c}u_0 , \quad \tilde{\rho}_1(\sigma,\chi) = \frac{\rho_0}{c}\left(u_1 + \frac{1}{c}u_0^{\ 2} + \frac{1}{c}\partial_\chi^{\ -1}u_{0,\sigma}\right) \\
&\tilde{\rho}_2(\sigma,\chi) = \frac{\rho_0}{c}\left(\begin{array}{l} u_2 + \dfrac{1}{c^2}u_0^{\ 3} + \dfrac{2}{c}u_0 u_1 + \dfrac{1}{c}\partial_\chi^{\ -1}u_{1,\sigma} + \dfrac{1}{c^2}\partial_\chi^{\ -2}u_{0,\sigma\sigma} \\ + \dfrac{1}{c^2}\partial_\chi^{\ -1}\left(u_{0,\chi}\partial_\chi^{\ -1}u_{0,\sigma}\right) + \dfrac{3}{c^2}\partial_\chi^{\ -1}\left(u_0 u_{0,\sigma}\right) \end{array}\right) \ .
\end{aligned} \tag{40}$$

Rescaling according to

$$u_0(\sigma,\chi) = c\,w(t.x), \quad u_1(\sigma,\chi) = c\,w_1(t.x), \quad \left(t = \frac{c^2(1+\gamma)^2\rho_0}{8\mu}\sigma \, , x = -\frac{c(1+\gamma)\rho_0}{2\mu}\chi\right) \ , \tag{41}$$

Eq. (38) becomes the perturbed Burgers equation for $w$, with a first-order perturbation, still containing $w_1$. One adopts for $w_1$ the differential polynomial in $w$ allowed by the formalism [16, 17]:

$$w_1(t,x) = a w^2 + b w_x \quad . \tag{42}$$

Eq. (39) now obtains the form of Eq. (32), with

$$\alpha_1 = \tfrac{2}{3} a\,, \quad \alpha_2 = -\tfrac{1}{3}\,, \quad \alpha_3 = \tfrac{2}{3} a + \tfrac{1}{4} - \tfrac{1}{12}\gamma\,, \quad \alpha_4 = \tfrac{1}{8} + \tfrac{1}{8}\gamma \quad . \tag{43}$$

Eq. (33), the condition for the absence of an obstacle, is not obeyed for these coefficients. (The formalism allows for a non-local term, $w_x\,\partial_x^{-1} w$. It is not included in $w_1$, because it is unbounded for multiple-wave solutions of the Burgers equation [27], and its inclusion does not alter the last conclusion. Also, the $w_x$ term in Eq. (43) does not contribute because it is a symmetry.)

Thus, it is impossible eliminate the obstacle if $w_1$ is a differential polynomial. However, if a non-polynomial term is allowed, the obstacle can be removed in the case of wave solutions of the Burgers equation. We modify Eq. (42) into

$$w_1(t,x) = a w^2 + b w_x + \zeta(t,x) \quad . \tag{45}$$

Instead of Eq. (32), one obtains

$$w_t = 2 w w_x + w_{xx} + \varepsilon\left(3\alpha_1 w^2 w_x + 3\alpha_2 w w_{xxx} + 3\alpha_3 w_x{}^2 + \alpha_4 w_{xxx} + 2(w\zeta)_x + \zeta_{xx} - \zeta_t\right) \quad . \tag{46}$$

The goal is to break Eq. (46), with the values of $\alpha_i$, $i$ = 1-4 of Eq. (43), into two equations: An asymptotically integrable perturbed Burgers equation for $w(t, x)$, and an equation for $\zeta(t, x)$ that will contain a driving term, for which $\zeta(t, x)$ is bounded. This is obtained by choosing

$$a = \tfrac{1}{8}(\gamma - 7) \quad . \tag{47}$$

and breaking Eq. (46) up into:

$$w_t = 2ww_x + w_{xx} + \varepsilon\,\delta\left(3w^2 w_x + 3ww_{xxx} + 3{w_x}^2 + w_{xxx}\right), \quad \left(\delta = \alpha_1 + \alpha_2 - \alpha_3 = \tfrac{1}{12}(\gamma - 7)\right) \tag{48}$$

which is an asymptotically integrable Normal Form [16, 17], and [27]

$$\begin{aligned} &\zeta_t = 2(w\zeta)_x + \zeta_{xx} + \kappa\,\partial_x(ww_x) + \omega\,\partial_x(2ww_x + w_{xx}) \\ &\quad\kappa = -\alpha_1 + 2\alpha_2 + \alpha_3 - 2\alpha_4 = -\tfrac{1}{3}(\gamma + 2) \\ &\quad\omega = -\alpha_1 - \alpha_2 + \alpha_3 + \alpha_4 = \tfrac{1}{24}(\gamma + 17) \end{aligned} \quad . \tag{49}$$

While there is no rigorous proof that the solution of Eq. (49) is always bounded, there is numerical evidence that the two driving terms in the equation generate a bounded $\zeta(t, x)$ when $w$ is a two-wave solution (i.e., three fronts) of the Burgers equation [27]. This implies that, at least for wave solutions of the Burgers equation, the obstacle to integrability may be removed.

In summary, it has been shown that the emergence of obstacles to integrability in perturbed evolution equations may be a consequence of the expansion procedure employed in their derivation from the equations of complex dynamical systems. Algorithms exist that lead to equations that are devoid of obstacles. Although the analysis has been carried out only through the first order, in which an obstacle emerges, the pattern for extension to higher orders is rather obvious. These results do, certainly, not preclude the possibility of the existence of obstacles to asymptotic integrability that have a sound physical basis.

Acknowledgment: A critical comment by A. Fokas is deeply acknowledged.

REFERNCES